\documentclass[
 aps,
 prl,
 showpacs,
 amsmath,
 amssymb,
 showkeys,
 reprint,%
]{revtex4-1}

\usepackage{graphicx}
\usepackage{dcolumn}
\usepackage{bm}
\usepackage{amsmath} 
\usepackage{hyperref}
\usepackage{breakurl}

\begin{document}

\preprint{AIP/123-QED}

\title[Hardware-based Demonstration of Time-delay Interferometry and TDI-ranging with Spacecraft Motion Effects]{Hardware-based Demonstration of Time-Delay Interferometry \\and TDI-Ranging with Spacecraft Motion Effects}

\author{Shawn J. Mitryk}
 \email{smitryk@phys.ufl.edu.}
 \author{Josep Sanjuan}
 \author{Guido Mueller}
\affiliation{Department of Physics, University of Florida, \\
PO Box 118440, Gainesville, FL 32611-8440, USA}

\date{\today}

\begin{abstract}
This hardware-based experimental simulation of the Laser Interferometer Space Antenna (LISA) implements a real-time electronic duplication of the time-changing inter-spacecraft (SC) laser phase delays while measuring heterodyned laser fields with $\mu$cycle phasemeters. The beatnote measurements are used to verify the capabilities of theorized post-processing time-delay interferometry (TDI) combinations in the proper time-scaled and time-shifted linear combinations. The experiments meet the 18\,pm$/\sqrt{\rm{Hz}}$ LISA measurement sensitivity goal after demonstrating the cancellation of 100\,Hz$/\sqrt{\rm{Hz}}$ laser phase noise by TDI-ranging the time-varying arm-length to an accuracy better than 2.0\,m using a frequency modulated ranging tone.
\end{abstract}

\maketitle

\section{\label{sec:Intro}Introduction}

Future space-based gravitational wave (GW) interferometers \cite{SpaceBasedRFI}, such as the Laser Interferometer Space Antenna (LISA) \cite{LISAPrePhaseA, LISAYellowBook}, will measure gravitational radiation from compact-star binaries and binary black hole mergers in the 0.1\,mHz to 0.1\,Hz frequency range, providing a new window through which to observe these astrophysical systems \cite{LISAScienceReq}. LISA exploits a modified Michelson GW detection technique by taking one-way laser phase measurements between laser benches on three individual spacecraft (SC) (Fig.\,\ref{fig:LISAConstModel} \cite{LISADesign}) to measure and extract the GW spacetime strain. The SC, defining the vertices of a triangular constellation, follow independent heliocentric orbits resulting in unequal, time-changing interferometer arm-lengths. Thus, the GW measurement sensitivity depends heavily on the ability to combine these individual SC data-sets to account for the laser phase noise, clock noise, and spacecraft motion. 
These time-scaled and time-shifted linear combinations, referred to as time delay interferometry (TDI) combinations \cite{TDIConcept}, complete the laser transfer chain, cancel the laser phase noise, and extract the GW strain information. LISA Simulator \cite{LISASimulator}, Synthetic LISA \cite{SynthLISA}, and LISA Tools \cite{LISATools} have produced numeric simulations of these data-sets for mock LISA data challenges (MLDCs) \cite{LISAMockData}. Hardware based laboratory experiments have also verified Sagnac-type TDI-combinations with clock noise corrections using a 1\,meter test-bed. \cite{SagnacTDI}

Taking the next step in validating the effectiveness of the TDI combinations, the University of Florida has constructed a hardware-based LISA simulator which utilizes a real-time digital electronic replication of the multi-second laser phase delays between individual laser benchtops while simultaneously taking low-frequency phasemeter (PM) measurements of LISA-like photodetector (PD) beatnotes \cite{RachaelUFLIS}. Previous experiments have generated data-sets and tested the capabilities of the TDI-$X_1$ combinations \cite{UFLIStdi1} which cancel the laser phase noise in a static interferometer. Advancements to the simulator have provided the ability to model time-changing delays and incorporate the SC-motion induced laser phase coupling into the measurements. The following experiments use the TDI-$X_2$ combinations to cancel $\simeq100$\,Hz$/\sqrt{\rm{Hz}}$ laser frequency noise in a LISA-like interferometer resulting in greater than 10 orders of magnitude noise suppression below 1\,mHz and meeting the 
18\,
pm$/\sqrt{\rm{Hz}}$ LISA measurement sensitivity goal in a majority of LISA-like experiments. The analysis also shows how the time-varying inter-SC round-trip arm-lengths are ranged to an accuracy of $<2.0$\,meters utilizing a TDI-ranging \cite{TDIRanging} reference tone.

\begin{table}
	\caption{\label{tab:Requirements} LISA Characteristics $\&$ Requirements \footnote{Note that LISA, as a combined NASA/ESA funded mission, no longer exists and has since been replaced by NGO/eLISA in Europe while NASA develops new space-based interferometer concepts under the acronym SGO. However, most space-based GW interferometers will be generally LISA-like with similar measurement complications. \cite{NGOSummary, SpaceBasedRFI, Lagrange, GADFLI} The requirements specified and experiments performed in this description focus on an approximate `worst-case' LISA-like scenario.}}
	\begin{ruledtabular}
		\begin{tabular}{cc}
			Characteristic & Specification\\
			\hline
			Laser Stab. & $\frac{280\,Hz}{\sqrt{Hz}} \sqrt{1+\left(\frac{f_M}{f}\right)^4}$ \\
 			PM Precision & $\frac{1\,\mu cycle}{\sqrt{Hz}} \sqrt{1+\left(\frac{f_M}{f}\right)^4} $ \\
			IMS Strain Sens.\footnote{The IMS strain sensitivity refers to a single link requirement including shot-noise, path-length noise, residual laser phase noise, phasemeter noise, and many other technical noise sources.  The experimentally relevant terms are used to calculate the TDI-$X_2$ displacement equivalent sensitivity goal.} & $\frac{18\,pm}{\sqrt{Hz}} \sqrt{1+\left(\frac{f_M}{f}\right)^4}$ \\
			DRS Accel. Noise & $\frac{3\,fm/s^{2}}{\sqrt{Hz}} \sqrt{1+\left(\frac{f}{f_H}\right)^4}\sqrt{1+\left(\frac{f_L}{f}\right)}$ \\
			\hline
			Ranging Accuracy & $\delta L = 1\,\rm{meter},~~~\delta\tau = 3.3\,$ns \\
			Arm-Length & $L = 5.0 \pm 0.1~Gm$ \\
			Light-Travel Delay & $\tau = 16.67 \pm 0.33~s$ \\
			Relative Velocity & $v = \pm 20~m/s,~~~\beta = \pm 66~ns/s$ \\
		\end{tabular}
		$f_L = 0.1$\,mHz, $f_M = 2.8$\,mHz, $f_H = 8$\,mHz \\
	\end{ruledtabular}
\end{table}

\section{\label{sec:ExpModel}Modeling LISA}

The sensitivity of the LISA detector is determined through a combination of requirements \cite{LISADesign} which are defined to optimally observe scientifically interesting astrophysical sources while staying within the bounds of cost and feasibility. Each element in the LISA measurement chain has a pre-defined requirement (Table\,\ref{tab:Requirements}) in order to meet the overall measurement sensitivity. Based on these specifications, the dominant sensitivity-limiting terms in the LISA design are the disturbance reduction system's (DRS) acceleration noise at low frequencies, $f<3$\,mHz, and the interferometry measurement system's (IMS) displacement sensitivity at high frequencies, $f>3$\,mHz. The DRS is implemented to limit non-gravitational acceleration noise on the six, gravitationally sensitive, proof masses. This will be verified with a pre-LISA test mission, the LISA-Pathfinder \cite{LISAPathfinder}. The IMS is responsible for measuring the one-way differential length between these proof masses.

\begin{figure}
\includegraphics[width=.48\textwidth]{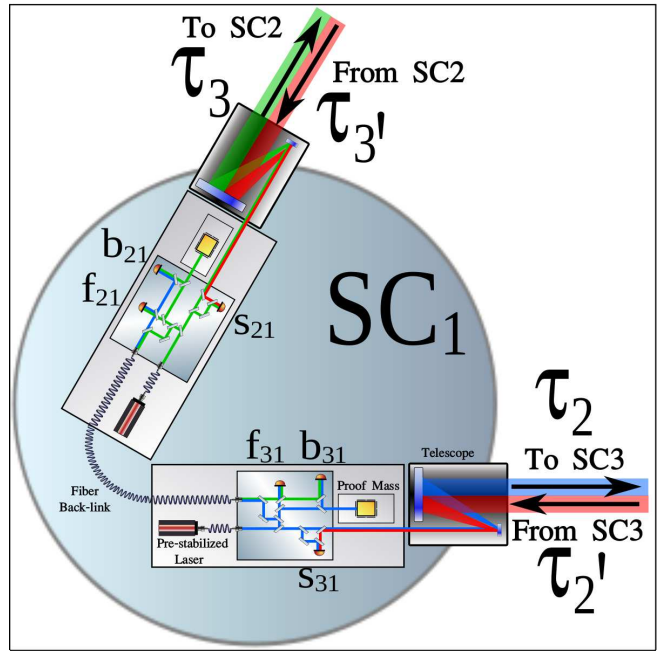}
\caption{\label{fig:LISAConstModel} Model of the LISA-IMS: The pre-stabilized lasers, proof-masses, optical benches, and inter-SC light field transfers of the LISA mission are illustrated. The spacecraft, $SC_{i}$, the associated light-travel time delays between the SC, $\tau_{q/q'}$, and the photodetector observables, $s_{sr}$, $b_{sr}$, and, $f_{sr}$, are labeled.}
\end{figure}

The LISA interferometry measurement scheme consists of six nearly-identical laser benchtops, two on each of the three SC. Each benchtop includes a pre-stabilized laser source, an optical bench, a proof-mass, a DRS, a fiber coupler to transfer the laser field between adjacent benchtops, and a telescope to transmit the laser field to the adjacent SC (Fig.\,\ref{fig:LISAConstModel}) \cite{LISADesign}. Each optical bench uses $\mu$cycle PMs to measure the differential laser phase of heterodyned laser fields on three primary PDs. The measured signals are encoded with the local-SC to far-SC distance, $s_{sr}$, the local-SC to local-proof-mass distance, $b_{sr}$, and the differential laser phase induced by the fiber back-link between adjacent benchtops, $f_{sr}$. The TDI combinations of these observables are derived to complete the phase transfer chain and cancel the dominant laser phase noise.

To clarify the TDI analysis we consider two simplifications to the system. First, we assume that the counter-propagating fiber back-link terms, $f_{sr}$, can be measured and accounted for and, thus, we may treat the SC as having only one laser source \cite{FiberNoise}. Also, assuming the DRS system works well enough to shield the proof-masses from non-gravitational acceleration noise sources and that the local SC to local proof-mass distance can be accurately measured and accounted for, it is then justifiable to interpret the SC themselves as the interferometric GW proof masses and the $b_{sr}$ terms can be neglected in the analysis as well \cite{LISAPathfinder}. This leaves the interesting TDI terms,
\begin{align}
 \label{eqn:phaseTransfer}
 s_{sr}(t) =& \phi_{r}(t) - \phi_{s}(\alpha_q (t- \tau_{q}(0))) + g_q(t), \\
 \label{eqn:SsrSignal}
 s_{sr} =& \phi_{r} - \phi_{s;q} + g_q,
\end{align}
or effectively, a comparative one-way measure of the local SC to far SC distance with a first-order velocity correction. In this notation $\phi_{s}$ is the phase of the `sending' laser (from the adjacent SC), $\phi_{r}$ is the phase of the `receiving' laser (on the local SC), and $g_q$ is the GW modulated laser phase on the arm opposite $SC_q$. The light-travel time-delays between the space-craft can written as $\tau_{q}(t) = \tau_{q}(0)+\beta_q t$ where $\tau_{q}(0)$ is the initial light-travel time-delay along the arm opposite $SC_q$, from $SC_s$ to $SC_r$. $\beta_{q} = [1-\alpha_{q}] = v_{q}/c$ where $v_{q}$ is the differential position, or velocity, between $SC_{r}$ and $SC_{s}$. The colon notation is used to transfer laser fields between moving frames by taking the time transformation, $t\rightarrow\alpha_q(t-\tau_q(0))$, as shown in Eq.\,\ref{eqn:SsrSignal}, of which can be applied successively as in Eq.\,\ref{eqn:SingleArm} \cite{ShaddockTDI2a}. Note that the counter-propagating inter-SC light-travel 
time-delays along a single arm are not equal ($\tau_{q}(0) \neq \tau_{q'}(0)$) due to the orbital rotation of the constellation, although in this analysis, the first derivative is: $\beta_{q} = \beta_{q'}$ \cite{ShaddockTDI2b}. The prime notation refers to the different outgoing (un-primed) and returning (primed) light-travel-time laser phase delays. It is assumed that $|v_{q}| < 300$\,m/s, or, $|\beta| < 10^{-6}$; thus, 2$^{nd}$ order special relativity corrections of order $\beta^2$ are negligible \cite{ShaddockTDI2a, CornishTDI}.

\subsection{TDI Theory}

Choosing a master SC, $SC_1$, as the interferometer vertex (beam-splitter), we can cancel the delayed $SC_{2/3}$ laser phase terms from the local, $s_{s1}$, signals and construct two differential, round-trip single-arm measurements by forming,
\begin{align}\label{eqn:Sensor}
 \Delta_{s} =& s_{s1} + s_{1s;q'}\\
\nonumber
 \Delta_{s} =& \phi_{1} - \phi_{s;q'} + (\phi_{s} - \phi_{1;q})_{;q'}\\
\nonumber
 \Delta_{s} =& \phi_{1} - \phi_{s;q'} + \phi_{s;q'} - \phi_{1;qq'}\\
 \label{eqn:SingleArm}
 \Delta_{s} =& \phi_{1} - \phi_{1;qq'}
\end{align}
or, explicitly as a function of time:
\begin{align}
 \Delta_{s}(t) =& s_{s1}(t) + s_{1s}(\alpha_{q'} (t-\tau_{q'}(0)))\\
\nonumber
 \Delta_{s}(t) =& \phi_{1}(t) + \phi_{1}(\alpha_{q}^2t - \alpha_{q}^2\tau_{q'}(0) - \alpha_{q}\tau_{q}(0))
\end{align}
where $\phi_{1}$ is the master pre-stabilized laser phase.

In the special case where the total round trip delay-times are equal, $[\tau_{2}+\tau_{2'}] = [\tau_{3}+\tau_{3'}]$, and the differential SC velocities are zero, $\beta_2 = \beta_3 = 0$, the difference of the sensor signals,
\begin{align}
 X_0 = \Delta_{2} - \Delta_{3},
\end{align}
generates a standard equal-arm Michelson interferometer output, independent of laser phase noise.\footnote{We diverge from previously published nomenclature and standardize the notation in this description to simplify the TDI expansion  and to account for variations in the inter-SC distance as outlined in Table.\,\ref{tab:TDIGenerations}} For LISA, this is rarely a reasonable laser phase cancellation technique since the arm-lengths are almost always un-equal. However, the TDI-$X_{1}$ combination \cite{ShaddockTDI1}, written as
\begin{align} \label{eqn:TDI1Comb}
 X_{1} = \Delta_{2} - \Delta_{3} -\Delta_{2;22'} + \Delta_{3;33'},
\end{align}
replicates an equal-arm interferometer phase delay and cancels the common laser phase noise in the case where $[\tau_{2}+\tau_{2'}] \neq [\tau_{3}+\tau_{3'}]$ and $\beta_2 - \beta_3 \simeq 0$. Calculating the timing-error between the $\phi_{1;33'22'}$ and $\phi_{1;22'33'}$ terms in Eq.\,\ref{eqn:TDI1Comb}, which result from laser phase transformation order of the $\Delta_{2;22'}$ and $\Delta_{3;33'}$ terms, Eq.\,\ref{eqn:phaseTransfer}, we find:
\begin{align}
 \delta \tau = 4 \tau |\beta_2 - \beta_3|
\end{align}
where $\tau$ is the mean one-way delay time ($\simeq$\,16.7\,s). Exploiting Eq.\,\ref{eqn:TDIXErr}, we can calculate the suppression limit of the TDI-$X_1$ combination, which fails to account for this SC-motion induced timing error, as \cite{CornishTDI}:
\begin{align} \label{eqn:TDICornish}
 \tilde{X}_{1} > 4\tau|\beta_2-\beta_3|\dot{\tilde{\phi}}_1
\end{align}
where $\dot{\tilde{\phi}}_1$ is the time-differentiated laser phase spectrum. Given a situation where this TDI-$X_{1}$ limit is large enough to restrain the IMS sensitivity, the TDI-$X_{2}$ combination,
\begin{align} \label{eqn:TDI2Comb}
 X_{2} =& \Delta_{2} - \Delta_{3} -\Delta_{2;22'} + \Delta_{3;33'} \\
 \nonumber
	 &~~~~- \Delta_{2;33'22'} + \Delta_{3;22'33'} \\
 \nonumber
	 &~~~~~~~~~~~~ + \Delta_{2;22'22'33'} - \Delta_{3;33'33'22'}
\end{align}
must be used to cancel the velocity coupled laser phase noise \cite{ShaddockTDI2a}. This TDI-$X_{2}$ combination accounts for the timing error of the TDI-$X_{1}$ combination by re-tracing and applying the TDI-$X_{1}$ laser phase-delay transfer chain through the constellation a second time.

The TDI-$X_1$ and TDI-$X_2$ combinations include multiple single-link, $s_{sr}$, signals. Thus, the allowed noise for these TDI-$X_1$ and TDI-$X_2$ increase by a factor of 2 and 4 respectively as compared to those referenced and accounted for in Table.\,\ref{tab:Requirements}.

\begin{table*}
    \caption{\label{tab:TDIGenerations} TDI generations based on orbital dynamics approximations \cite{ShaddockTDI2a, SynthLISA}}
    
    \begin{tabular}{lccc}
    \hline
    Generation$~~~~$ & $~~~~$Michelson Arm-Length$~~~~$ & $~~~~$Counter-Propagating Delay$~~~~$ & $~~~~$Delay Dynamics$~~~~$\\
    \hline
    TDI-$X_{0.0}$  & $\tau_{22'}(t)  =   \tau_{3'3}(t)$ &  $\tau_{q}(0) = \tau_{q'}(0)$ & $d\tau_{q}(t)/dt = 0$ \\
    TDI-$X_{1.0}$ (First Generation TDI)  & $\tau_{22'}(t) \neq \tau_{3'3}(t)$ &  $\tau_{q}(0) = \tau_{q'}(0)$ & $d\tau_{q}(t)/dt = 0$ \\
    TDI-$X_{1.5}$ (Modified TDI) & $\tau_{22'}(t) \neq \tau_{3'3}(t)$ &  $\tau_{q}(0) \neq \tau_{q'}(0)$ & $d\tau_{q}(t)/dt = 0$ \\
    TDI-$X_{2.0}$ (Second Generation TDI) & $\tau_{22'}(t) \neq \tau_{3'3}(t)$ &  $\tau_{q}(0) \neq \tau_{q'}(0)$ & $d\tau_{q}(t)/dt = \beta_{q}$ \\
    TDI-$X_{3.0+}$  & $\tau_{22'}(t) \neq \tau_{3'3}(t)$ &  $\tau_{q}(0) \neq \tau_{q'}(0)$ & $d\tau_{q}(t)/dt = \beta_{q}(t)$ \\ 
    \hline
    \end{tabular}

\end{table*}

Although the following analysis will focus on the TDI-$X_{2}$ velocity corrections, annual changes in $\beta$, or SC acceleration terms, may be accounted for with further expansion of these TDI combinations. Otherwise and in order to utilize the TDI-ranging methods outlined below for LISA TDI data-analysis, the $\beta$ value will have to be adjusted to avoid the acceleration-dependent accumulated error. Even though a continuous measure and correction to the $\beta$ values is possible, these adjustments may also be accomplished by segmenting the data-analysis and adjusting the ranging functions used to evaluate the TDI combinations at regular intervals, in a worst case LISA-like scenario, every $\sqrt{\delta\tau~ T_{year}/(\beta~\pi)} = \sqrt{3.3\,\rm{ns}*3.15\times10^{7}\,\rm{s}/(66\,\rm{ns/s}*\pi)} =$ 708\,s \cite{SynthLISA}.

\begin{figure*}
\includegraphics[width=.95\textwidth]{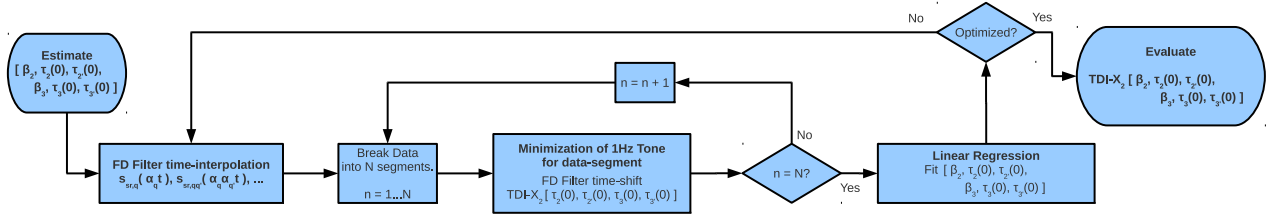}
\caption{\label{fig:FlowChart} Flow chart of the ranging-tone minimization process: This process minimizes the ranging tone and maximally constrains the six variable light travel time delays resulting in an optimized TDI-$X_2$ strain sensitivity. The results of this process for the different experimental configurations are in Table\,\ref{tab:RangingResults}. }
\end{figure*}

\begin{figure*}
\includegraphics[trim=2cm 0cm 2cm 0cm, clip=true, width=.95\textwidth]{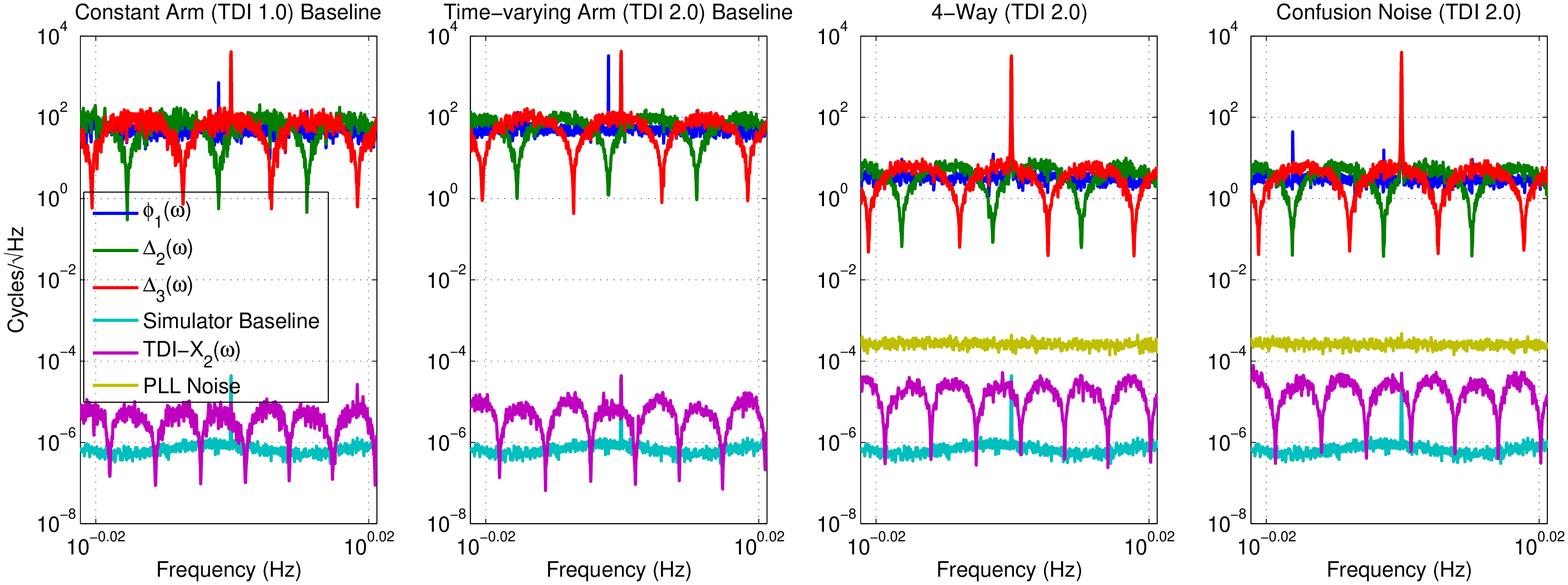}
\caption{\label{fig:TDIToneSuppression} Spectral magnitude tone suppression: The ranging tone modulated inputs are plotted along with the resulting TDI-$X_2$ combinations for the four different experiments outlined in Table\,\ref{tab:TDIMeasurements}. The decreased round-trip timing accuracy of the $\Delta_2$ arm as compared to the $\Delta_3$ arm as shown in Table\,\ref{tab:RangingResults} is likely due to the proximity of the nearest arm-zeros.}
\end{figure*}

\subsection{TDI Ranging}
\label{sec:TDIRanging}
Until now, it has been assumed that we know the required interferometer arm-lengths and rate of change in order to form the TDI combinations but in practice this is not the case. Two primary methods, pseudo-random noise (PRN) ranging \cite{PRNRanging, PRNRanging2} and TDI-ranging \cite{TDIRanging}, have been proposed to measure these time-dependent arm-lengths. Extending the root mean squared (RMS) power minimization TDI-ranging methods outlined by Tinto in \cite{TDIRanging}, this experiment will, instead, modulate the laser phase signals with ranging reference tones at frequencies outside of the LISA measurement band. The RMS power minimization around these relatively high-frequency tones avoids the displacement of the measured arm-lengths as a result of low-frequency GW signals \cite{TDIRanging} and provides an improved ranging precision beyond the inherent RMS laser noise cancellation resulting from the larger signal power at these chosen frequencies.

Using the Taylor approximation,
\begin{align}
 \nonumber
 X_{Err}(t) \simeq&~~\phi(t) - \phi(t+\delta\tau), \\
 \nonumber
 \tilde{X}_{Err}(\omega) \simeq&~~[e^{-i\omega t} - e^{-i\omega (t + \delta\tau)}]\tilde{\phi}, \\
 \label{eqn:TDIXErr}
 |\tilde{X}_{Err}(\omega)| \simeq&~~\omega\delta\tau |\tilde{\phi}|,
\end{align}
for $\omega\delta\tau << 1$, we can estimate a simplified but reasonable measure of the arm-lengths through the cancellation of these ranging tones using the TDI combinations to an accuracy of:
\begin{align} \label{eqn:RangingAcc}
 \delta L = \delta\tau c \simeq \frac{c}{2\pi f_{Tone}}G_{Tone},
\end{align}
where $f_{Tone}$ is the ranging-tone modulation frequency (1\,Hz for these experiments) and $G_{Tone}$ is the tone suppression magnitude when evaluated with the TDI combination. Generally, the cancellation of the local laser signal's ranging tones, $\phi_1(t)$, from the far $s_{sr}$ signals in the TDI-X combinations (Eq.\,\ref{eqn:TDI1Comb}, \ref{eqn:TDI2Comb}) constrains the one-way outgoing delay times, $\tau_{2}(t)$ and $\tau_{3}(t)$, while the cancellation of the far laser signal's tones, $\phi_2(t)$ and $\phi_3(t)$, from the local $s_{sr}$ signals constrains the incoming delay times, $\tau_{3'}(t)$ and $\tau_{2'}(t)$, respectively. Exploiting the phase-locking methods outlined in \cite{PLLLocking} such that $s_{12} \simeq s_{13} \simeq 0$ or, equivalently, $\phi_{2} = \phi_{1;3'}$ and $\phi_{3} = \phi_{1;2'}$, we can transfer the local laser phase data and phase stability to the far lasers and constrain the round trip delay times:
\begin{align}\
	\label{eqn:RoundTripDelay2}
	\tau_{22'}(t) = \alpha_{2}^2t - \alpha_{2}^2\tau_{2'}(0) - \alpha_{2}\tau_{2}(0), \\
	\label{eqn:RoundTripDelay3}
	\tau_{33'}(t) = \alpha_{3}^2t - \alpha_{3}^2\tau_{3'}(0) - \alpha_{2}\tau_{3}(0),
\end{align}
using a local ranging tone only.

Although some prior estimate of the arm-lengths will likely exist, the analysis in this paper will assume no previous knowledge of the 6-variable time-dependent round-trip arm-lengths, $\tau_{22'}(t)$ and $\tau_{33'}(t)$, and will determine these arm length functions using a 1\,Hz laser frequency modulation with an amplitude of 500\,Hz. Note that the ranging tone should not be placed at a frequency which is near an integer multiple of the interferometer arms' inverse round-trip delay time to avoid the inherent tone cancellation along a single arm as shown in Fig.\,\ref{fig:TDIToneSuppression}. This local ranging tone will provide constraints on the round trip delay times only. Since the far laser phase signals in this experiment are not modulated with ranging tones but rather, phase locked \cite{PLLLocking}, the one-way delay times are constrained using the TDI-ranging methods outlined by Tinto \cite{TDIRanging} through the minimization of the phase lock loops's (PLL) residual phase RMS power in the TDI-$X_{
2}$ combination.

As shown in Fig.\,\ref{fig:FlowChart}, initially assuming the arm-lengths are constant ($\beta = 0$) a four-dimensional sweep of the `time-space' defined by $\tau_{2}(0), \tau_{2'}(0), \tau_{3}(0),$ and $\tau_{3'}(0)$ is performed using a 51-point Lagrange fractional delay filter \cite{FDFilter} which determines the values which minimize RMS power near the 1\,Hz frequency-modulated ranging tone in the TDI-$X_2$ combination. Generally, depending on the pre-stabilized laser noise and phase locking configuration, the delay times must be measured with a $10^{-8}$ resolution \cite{LISADesign, PLLLocking}. Therefore, to make data analysis more efficient, rather than evaluating the $10^8$ possible values along each of the 4 delay-dimensions, the time-delay determination is performed using an algorithmic scan with successively finer time-delay mesh grids until the ranging tone is minimized and dominated by the noise floor of the IMS measurement. 

Using this scanning process, we determine an initial measure of the round-trip delay offsets, $\tau_{22'}(0)$ and $ \tau_{33'}(0)$, which minimize the ranging tone for small data-segments along a continuous data-set. The fitted slope of these offsets evaluates a first-order approximation of $\beta$ and the time-dependent arm-lengths. Applying the $\beta$-value dependent time-expansion or time-contraction to the $s_{sr}$ signals using time-varying fractional delay interpolation \cite{FDFilter}, the process is repeated iteratively, further optimizing the arm-length functions as shown in Fig.\,\ref{fig:FlowChart}. Finally, the fitted values, $\beta_2, \beta_3, \tau_{2}(0), \tau_{2'}(0), \tau_{3}(0),$ and $\tau_{3'}(0)$, are used to evaluate the TDI-$X_2$ combination along the entire data-set. The results produce a measure of the round-trip arm-lengths and place constraints on the one-way arm-lengths to an accuracy beyond the ranging requirements, thus, removing the sensitivity-limiting laser and PLL phase 
noises sources and 
idealizing the total interferometer strain precision.

\begin{figure*}
\includegraphics[width=.8\textwidth]{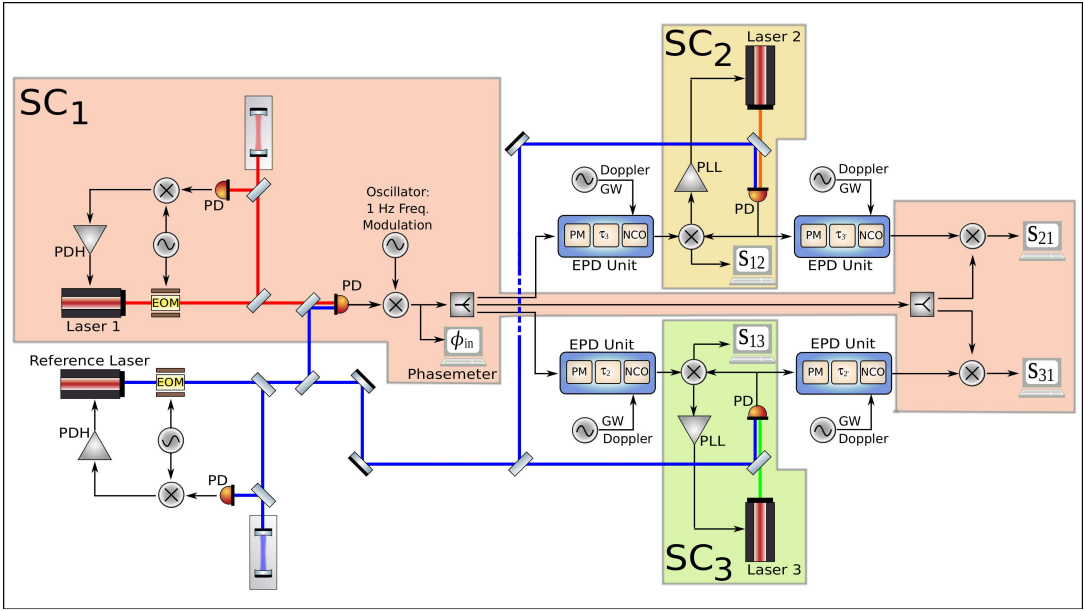}
\caption{\label{fig:TDIExperiment} Experimental Model of the LISA Interferometry Benchtop: The arrangement of the laser phase sources, EPD units, phase-lock loops, PD signal beatnotes, and PM measurements which are used to emulate the LISA-IMS in our experiments are presented. The measured $s_{sr}$ signals are used to form the TDI combinations.}
\end{figure*}

\section{LISA Simulator Benchtop}
The University of Florida LISA Simulator (UFLIS) benchtop (Fig.\,\ref{fig:TDIExperiment}) consists of four lasers, three of which, Laser-1 (L$_1$), Laser-2 (L$_2$), and Laser-3 (L$_3$) represent the lasers on each of the SC in the LISA constellation. Laser beatnotes are formed between each of these laser fields and a global reference laser (L$_{R}$), acting as an optical clock against which each individual laser phase is measured. Combinations of these PD measurements cancel the common L$_{R}$ phase noise. L$_{R}$ and L$_{1}$ are stabilized through Pound-Drever-Hall (PDH) stabilization \cite{PDHStabilization} using ultra-low expansion reference cavities to achieve a $\simeq100$ Hz/$\sqrt{\rm{Hz}}$ pre-stabilized laser noise as shown in Fig.\,\ref{fig:TDI1Rescaled}. Even though a lower laser frequency noise is achievable \cite{RachaelUFLIS}, it is intentionally spoiled to display the suppression capabilities of the TDI combinations.

\begin{table*}
	\caption{\label{tab:TDIMeasurements} TDI Experimental Characteristics: Four 40000\,s experiments are performed with increasingly more complicated, yet more LISA-like, characteristics. The static transponder experiment provides us with a baseline measure of the experimental setup's noise performance. The dynamic transponder experiment demonstrates the ability to determine and account for the time-changing delay-times. The phase-locked LISA-like experiment proves the ability to remove independent SC noise sources and constrain one-way delay times. Finally, the confusion noise experiment verifies that the TDI-ranging capability will not be limited by low-frequency LISA noise sources.}
	\begin{ruledtabular}
		\begin{tabular}{llll}
			Simulation Name & $s_{1r}$ & $\beta$ (ns/s) & Verification Signal\\
			\hline
			Static Transponder & $s_{1r}\simeq0$ & $\beta_2$ = $\beta_3$ = 0 & 6.22\,mHz Binary\\
			\hline
			Dynamic Transponder & $s_{1r}\simeq0$ & $\beta_2 = -100$, $\beta_3 = +150$ & 6.22\,mHz Binary\\
			\hline
			Dynamic LISA-like & $s_{1r}\simeq\phi_{PLLr}$ & $\beta_2 = -100$, $\beta_3 = +150$ & 6.22\,mHz Binary\\
			~ & ~~~~~$\simeq (1.0/f)\,\rm{mcycle}/\sqrt{\rm{Hz}}$ & ~ & ~ \\
			\hline
			Dynamic LISA-like & $s_{1r}\simeq\phi_{PLLr}$ & $\beta_2 = -100$, $\beta_3 = +150$ & 6.22\,mHz Binary\\
			~with Confusion-Noise & ~~~~~$\simeq (1.0/f)\,\rm{mcycle}/\sqrt{\rm{Hz}}$ & ~ & ~~+ Confusion-Noise \\
		\end{tabular}
	\end{ruledtabular}
\end{table*}

\subsection{Phasemeter}

The phasemeter is used to measure the phase of a 2-20\,MHz PD beatnote signal to an accuracy of $\simeq1\,\mu$cycle/$\sqrt{\rm{Hz}}$ (Table\,\ref{tab:Requirements}). The beatnote is sampled using a 14-bit analog-to-digital Converter (ADC) at a rate of 40\,MHz. The sampled signal is processed by a field programmable gate array (FPGA) programmed with a digital PLL. The digital PLL's feedback signal is down-sampled to a rate of 19.1\,Hz and relayed to a data-processing computer. Using differential and entangled phase \cite{MitrykDissertation, UFLIStdi1} PM measurements it has been determined that the PM is limited by $\tilde{\phi}_{ADC}$ (Fig.\ref{fig:TDI2Transponder}), a combination of frequency-dependent ADC timing jitter ($\delta t = 35/\sqrt{f}\,\rm{fs}/\sqrt{\rm{Hz}}$), RF-transformer phase dispersion, and amplitude noise \cite{MitrykDissertation}. One may write the PM measurement output as
\begin{align}
 \label{eqn:PMNoise}
 \phi_{PM}(t) =& \phi_{in}(t) + \frac{f_{in}}{f_{Clk}}\phi_{Clk}(t) + \phi_{ADC}(t),
\end{align}
where $\phi_{in}$ is the phase information on the $f_{in}$-frequency beat-note, $\phi_{Clk}$ is the phase noise of the $f_{Clk}$-frequency sampling clock, and $\phi_{ADC}$ are the ADC-noise sources mentioned above. Note the coupling of the clock's phase noise into the measurement. PM measurements taken on different LISA SC will be clocked using different ultra-stable clock sources requiring the need for clock-noise transfers between SC to remove these terms \cite{ClockNoise}. Although the following TDI experiments are taken use a single clock, the TDI combinations will still require clock-noise corrections to account for the time-shifted PM clock noise terms. This is discussed in the following section.

\subsection{\label{sec:EPDUnit}Electronic Phase Delay Unit}

The EPD unit simulates the characteristics of the laser phase transmission between the SC including the time-varying light travel time, Doppler shift, and gravitational wave phase modulations using a high-speed digital signal-processing (DSP) system. The front end is a fast PM, similar to the PM described above, with a data rate of 19.53\,kHz instead of 19.1\,Hz. The PM data is interpolated to time-lead or time-lag the phase information, producing a linear variation in the time-delay. GW signals and a Doppler offset are added to interpolated phase information before it is used to drive a numerically controlled oscillator (NCO). Completing the laser phase transmission replication, the NCO output is regenerated using a 16-bit digital-to-analog converter (DAC) with the same clock-source as the PM. After accounting for the clock noise coupling and Doppler shifts, we can write the EPD unit's output as,
\begin{align}
 \label{eqn:EPDNoise}
 \phi_{EPD}(t) =& \phi_{in}(t-\tau(t)) \\
\nonumber
 &~~+ \phi_{ADC}(t-\tau(t))+\phi_{DAC}(t) \\
\nonumber
 &~~+ \phi_{Clk:ADC}(t-\tau(t))-\phi_{Clk:DAC}(t) \\
\nonumber
 &~~+ \frac{f_{in}\pm f_{Dop}}{f_{Clk}}[\phi_{Clk}(t-\tau(t))-\phi_{Clk}(t)].
\end{align}
$\phi_{in}(t)$ is the phase evolution of the beatnote at an average frequency of $f_{in} = 2-20\,\rm{MHz}$. The beat signals are measured and regenerated with respect to a single clock, $\phi_{Clk}(t)$, at the DSP system clock frequency of $f_{Clk} = 40\,\rm{MHz}$. The phase noise of this clock enters as phase variations between the sampling and the Doppler shifted, $f_{Dop}$, regeneration after a time-varying time-delay $\tau(t) = \tau(0) + \beta t$. The single clock source is split and distributed between the sampling ADC and regeneration DAC causing an additional phase error, $\phi_{Clk:ADC}(t-\tau(t))-\phi_{Clk:DAC}(t)$. The sampling ($\phi_{ADC}$) and regeneration ($\phi_{DAC}$) processes add additional converter-noise contributions. It has been determined using two different sampling and regeneration clock sources that the limiting noise source on the EPD unit (Fig.\ref{fig:TDI2Transponder}) is the differential clock phase noise, $\phi_{Clk:ADC,\tau(t)}-\phi_{Clk:DAC}$, caused by errors in the clock 
distribution \cite{MitrykDissertation}. Note, as one may check, that the clock noise terms, $\phi_{Clk}(t)$ in Eq.\,\ref{eqn:EPDNoise}, themselves will cancel when measured with phasemeters (Eq.\,\ref{eqn:PMNoise}) and evaluated in the TDI combinations (Eq.\,\ref{eqn:TDI1Comb}, \ref{eqn:TDI2Comb}).

\subsection{Experimental Setup}

The laser benchtop, PMs, and EPD units are combined to recreate the LISA-IMS (Fig.\,\ref{fig:TDIExperiment}). The L$_1/$L$_{R}$ differential beatnote phase represents the pre-stabilized `input' noise. This PD signal is electronically mixed with a 1\,Hz frequency-modulated oscillator to add a ranging-tone and produce the input laser phase signals. Replicating an interferometer beam-splitter, these signals are electronically split and processed by the EPD units to simulate the outgoing-field inter-SC light transmission. The EPD-processed signal is mixed with each of the L$_{2/3}/$L$_{R}$ beatnotes `on the far spacecraft' to produce the $s_{12}$ and $s_{13}$ PM signals. These delayed signals are also used to phase-lock L$_{2/3}$, transferring the L$_1$ stability to these lasers \cite{PLLLocking}. The L$_{2/3}/$L$_{R}$ beatnotes are again processed by EPD units, simulating the returned-field inter-SC light transmission. Finally, the local differential L$_1/$L$_{R}$ input phase signal is mixed with the delayed L$_
{2/3}/$L$_{R}$ beatnotes to produce the $s_{21}$ and $s_{31}$ PM signals. It should be explicitly noted that these LISA-like $s_{sr}$ signals only include noise due to the pre-stabilized laser noise source, the phase-lock loop's phase noise, and EPD unit's ability to mimic the inter-SC time delay and are not sensitive to path-length differences or the shot-noise limit of the laser bench itself. Excluding these displacement and shot noises from the IMS noise budget reduces the single link requirement to about $5\,\rm{pm}/\sqrt{Hz}$, the TDI-$X_1$ combination requirement to $14.1\,\rm{pm}/\sqrt{Hz}$, and the TDI-$X_2$ combination requirement to $20\,\rm{pm}/\sqrt{Hz}$ \cite{LISADesign}.

Implementing this experimental model, four measurements are performed as outlined in Table\,\ref{tab:TDIMeasurements}. In the transponder measurements, rather than phase-locking the outgoing-field's EPD signal, it is transferred directly to the return-field's EPD unit as though it were reflected off of a moving mirror; accordingly, these TDI experiments may completely neglect the $s_{1r}$ terms while ranging is only required for the round-trip delay times, $\tau_{22'}(t)$ (Eq.\,\ref{eqn:RoundTripDelay2}) and $\tau_{33'}(t)$ (Eq.\,\ref{eqn:RoundTripDelay3}), instead of all four one-way delay times. For all measurements presented, the arm-lengths are chosen as $\tau_2 \simeq \tau_{2'} \simeq 16.55$\,s and $\tau_3 \simeq \tau_{3'} \simeq 16.75$\,s to maximize the unequal arm-length mismatch. The relative spacecraft velocities for the different measurements are shown in Table\,\ref{tab:TDIMeasurements}. They are artificially large to increase the $|\beta_2-\beta_3|$ limitations (Eq. \,\ref{eqn:TDICornish}) for 
the TDI-$X_1$ combination and to prove the ability of the TDI-$X_{2}$ to remove the differential velocity dependent noise from the TDI-$X_1$ combination. Doppler shifts of -2.0\,MHz and +3.0\,MHz are applied in order to produce the necessary MHz beatnote observables. A frequency modulated 6.22\,mHz verification binary GW signal with an amplitude of $\delta f_{GW} = 2\,\mu$Hz is injected into all four measurements to verify GW extraction. This frequency modulation equates to a phase modulation amplitude of $\delta\phi_{GW} = \delta f_{GW}/(2\pi f) = 51.2\,\mu$cycles resulting in a one-way strain amplitude of $\delta h = \delta \phi [\lambda /(c\tau)] = 1.1\times10^{-20}$. The resulting GW strain amplitude $h = 4\delta h = 4.4\times10^{-20}$ is a factor of 100 larger than expected for the RX J0806.3+1527 AM CVn binary \cite{LISAVerBinary}. Finally, a mock-confusion noise is added to the signals to demonstrate that this low-frequency noise will not limit the ranging capabilities.

The following experimental results are averaged over six, 10000\,s, data-segments during the course of a continuous 40000\,s experimental run-time.

\section{\label{sec:Results}TDI Results}

\begin{figure}
\includegraphics[trim=2cm 0.5cm 3cm .5cm, clip=true, width=.48\textwidth]{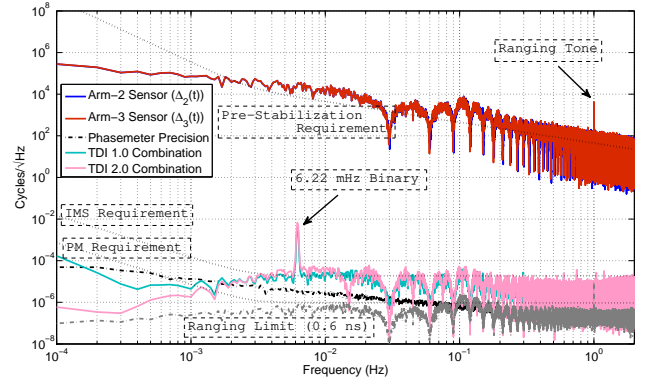}
\caption{\label{fig:TDI1Raw} Static Transponder (Baseline) Experimental Results: The sensor signals ($s_{21}(t) = \Delta_{2}(t)$, $s_{31}(t) = \Delta_{3}(t)$) are plotted along with the raw TDI-$X_1$ and TDI-$X_2$ results. The phasemeter measurement limitation and expected ranging limitations based on the calculated timing variance are also plotted.}
\end{figure}

\begin{figure}
\includegraphics[trim=2cm 0.5cm 3cm .5cm, clip=true, width=.48\textwidth]{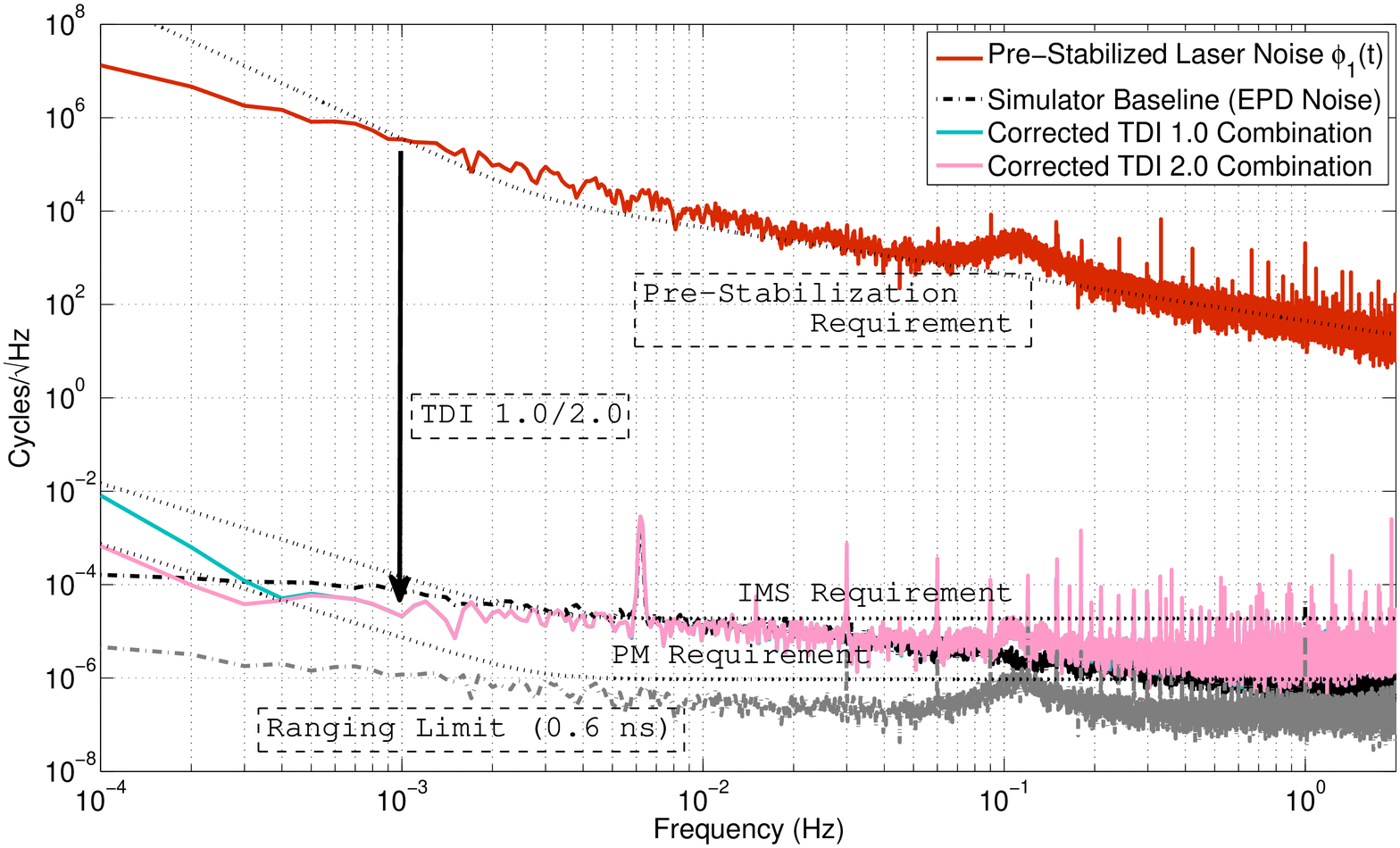}
\caption{\label{fig:TDI1Rescaled} Corrected Static Transponder (Baseline) Experimental Results: The input signal, $\phi_{1}(t)$, is plotted together with the TDI-$X_1$ and TDI-$X_2$ results, corrected by their respective transfer functions. Both TDI combinations are limited by the noise added by the EPD unit (grey curve).}
\end{figure}

\begin{figure}
\includegraphics[trim=2cm 0.5cm 3cm .5cm, clip=true, width=.48\textwidth]{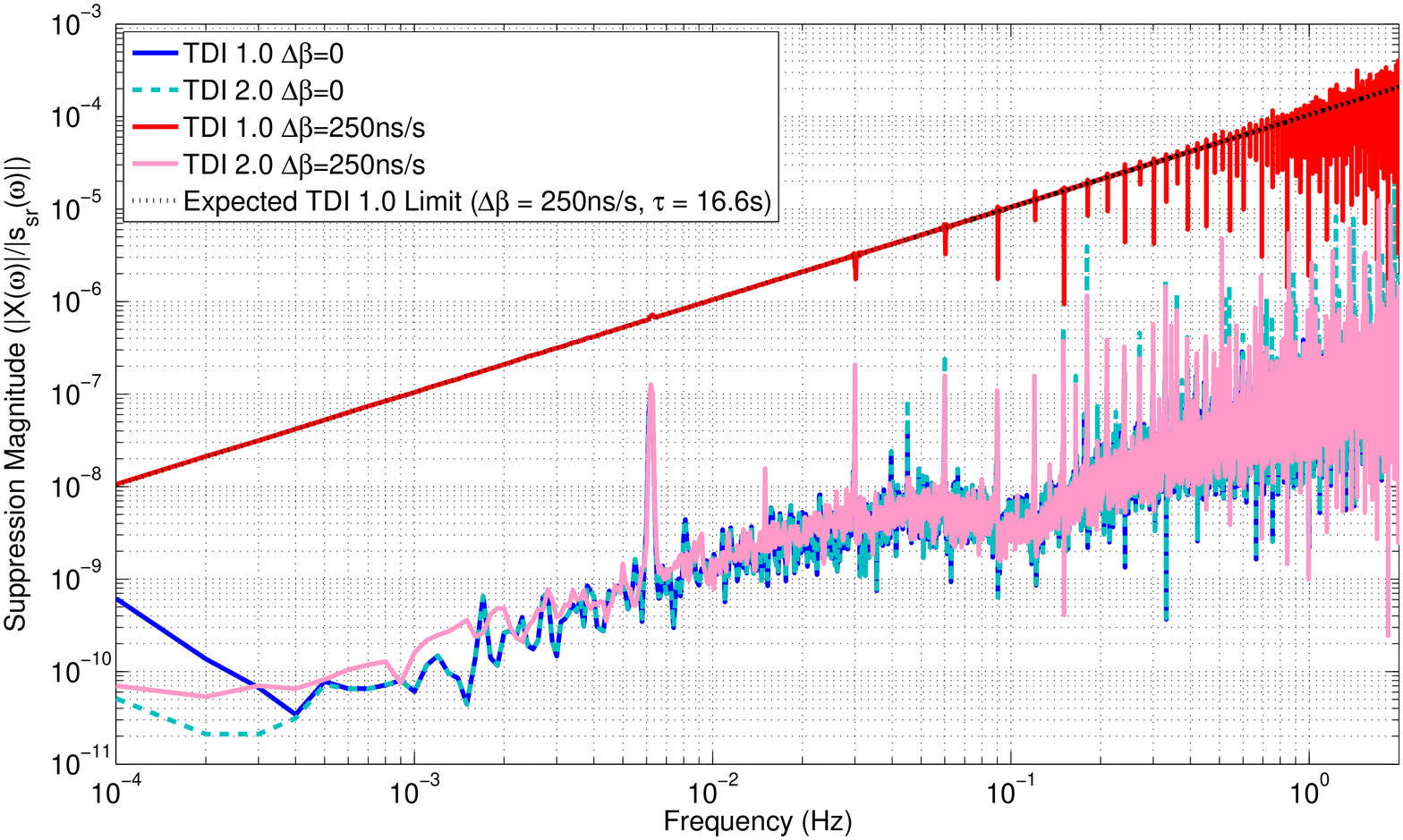}
\caption{\label{fig:TDISuppression} Laser Noise Magnitude Suppression Function: The achieved transponder-measurement laser noise suppression magnitudes of the TDI-$X_1$ and TDI-$X_2$ combinations are plotted for both the static and dynamic experiments. The TDI-$X_1$ combination's noise suppression equals the theorized limit (Eq.\,\ref{eqn:TDICornish})}
\end{figure}

\subsection{Static Arm Transponder \\ (Baseline)}
\label{sec:TDIStatic}
We first re-establish \cite{UFLIStdi1} a baseline ranging and measurement precision with static arm-lengths ($\beta=0$), utilizing the PD measurements of $s_{21}$ and $s_{31}$ and assuming $s_{12}, s_{13} \simeq 0$ when evaluating the TDI combinations. The 40000\,s data-set is broken into 40, 1000\,s segments. The first iteration and linear regression of the ranging process described in Fig.\,\ref{fig:FlowChart} produces a slope error (constraint on the arm-length velocities), of $|2\beta|<50$\,fs/s and a variance (round-trip ranging accuracy) of 0.6\,ns ($\sim$0.18\,m) as shown in Table\,\ref{tab:RangingResults}. In this experiment, we note that the TDI-$X_1$ combination's ranging-tone minimization produces the same result as the TDI-$X_2$ combination to within the measurement error. If $\beta\neq0$, this would not be the case since the ranging tone minimization using the TDI-$X_1$ combination would be limited by Eq.\,\ref{eqn:TDICornish} and would tend to calculate the mean delay for a particular data-
segment.

Using the calculated values we form the TDI-$X_1$ and TDI-$X_2$ combinations for the entire data-set. The raw TDI results, as plotted in Fig.\,\ref{fig:TDI1Raw}, show the laser noise cancellation and reveal the phase-modulated GW binary at 6.22\,mHz. The residual noise is dominated by phase noise added by the EPD units and, based on the timing-error, not by uncertainties in the ranging. The differences between the TDI-$X_1$ and TDI-$X_2$ combinations' spectral noise are caused by differences in their transfer functions with respect to the input laser phase noise. Fig.\,\ref{fig:TDI1Rescaled} shows the spectra after the TDI combinations have been rescaled by their respective transfer function magnitude. Both agree well with each-other and demonstrate greater than 10 orders of magnitude laser phase noise suppression below 1\,mHz. (Fig.\,\ref{fig:TDISuppression})

\begin{table*}
	\caption{\label{tab:RangingResults} Ranging Precision}
	\begin{ruledtabular}
		\begin{tabular}{llll}
			TDI Experiment Name & ~ & ~ & ~\\
			~ & ~ & ~ & ~\\
			~~~TDI-Ranging Constraint & ~ & ~ & ~\\
			\hline
			\hline
			Iteration (TDI combination) & $\beta$ & $\tau_{22'}(0), \tau_{33'}(0)$ & $\delta\tau_{22'}, \delta\tau_{33'}$ \\
			\hline
			\hline
			~ & ~ & ~ & ~ \\
			Static Transponder & ~ & ~ & ~ \\
			~ & ~ & ~ & ~\\
			~~~Round-trip Ranging & ~ & ~ & ~\\
			\hline
			\hline
			1 (TDI 1.0) & $2\beta_2 = -44.5$\,fs/s $\pm 20.9$\,fs/s & $\tau_{22'}(0) = 33.55204887148$\,s $\pm 0.23$\,ns & $\delta\tau_{22'} = 0.54$\,ns \\
			~ & $2\beta_3 = -46.3$\,fs/s $\pm 12.5$\,fs/s & $\tau_{33'}(0) = 33.15222859583$\,s $\pm 0.14$\,ns & $\delta\tau_{33'} = 0.32$\,ns \\
			\hline
			1 (TDI 2.0) & $2\beta_2 = -41.0$\,fs/s $\pm 21.2$\,fs/s & $\tau_{22'}(0) = 33.55204887151$\,s $\pm 0.24$\,ns & $\delta\tau_{22'} = 0.55$\,ns \\
			~ & $2\beta_3 = -46.3$\,fs/s $\pm 12.7$\,fs/s & $\tau_{33'}(0) = 33.15222859579$\,s $\pm 0.14$\,ns & $\delta\tau_{33'} = 0.33$\,ns \\
			\hline
			\hline
			~ & ~ & ~ & ~ \\
			Dynamic Transponder & ~ & ~ & ~ \\
			~ & ~ & ~ & ~\\
			~~~Round-trip Ranging & ~ & ~ & ~\\
			\hline
			\hline
			1 (TDI 2.0)& $2\beta_2 = -200.247$\,ns/s $\pm 100$\,ps/s & $\tau_{22'}(0) = 33.5518847$\,s $\pm 2.3\,\mu$s & $\delta\tau_{22'} = 7.5\,\mu$s \\
			~ & $2\beta_3 = +300.056$\,ns/s $\pm 95$\,ps/s & $\tau_{33'}(0) = 33.1525027$\,s $\pm 2.2\,\mu$s & $\delta\tau_{33'} = 7.0\,\mu$s\\
			\hline
			2 (TDI 2.0)& $2\beta_2 = -199.9998668$\,ns/s $\pm 80$\,fs/s & $\tau_{22'}(0) = 33.5519484187$\,s $\pm 1.8$\,ns & $\delta\tau_{22'} = 5.9$\,ns \\
			~ & $2\beta_3 = +300.0001130$\,ns/s $\pm 23$\,fs/s & $\tau_{33'}(0) = 33.1523897572$\,s $\pm 0.51$\,ns & $\delta\tau_{33'} = 1.7$\,ns \\
			\hline
			3 (TDI 2.0)& $2\beta_2 = -200.0000058$\,ns/s $\pm 8.9$\,fs/s & $\tau_{22'}(0) = 33.55194832884$\,s $\pm 0.20$\,ns & $\delta\tau_{22'} = 0.65$\,ns \\
			~ & $2\beta_3 = +300.0001361$\,ns/s $\pm 4.5$\,fs/s & $\tau_{33'}(0) = 33.15238977691$\,s $\pm 0.10$\,ns & $\delta\tau_{33'} = 0.33$\,ns \\
			\hline
			\hline
			~ & ~ & ~ & ~ \\
			Dynamic LISA-like & ~ & ~ & ~\\
			~ & ~ & ~ & ~\\
			~~~Round-trip Ranging & ~ & ~ & ~\\
			\hline
			\hline
			1 (TDI 2.0)& $2\beta_2 = -199.984$\,ns/s $\pm 12$\,ps/s & $\tau_{22'}(0) = 33.59821021$\,s $\pm 0.28\,\mu$s & $\delta\tau_{22'} = 0.895\,\mu$s \\
			~ & $2\beta_3 = +300.052$\,ns/s $\pm 7.8$\,ps/s & $\tau_{33'}(0) = 33.21476669$\,s $\pm 0.18\,\mu$s & $\delta\tau_{33'} = 0.568\,\mu$s\\
			\hline
			2 (TDI 2.0)& $2\beta_2 = -200.000015$\,ns/s $\pm 71$\,fs/s & $\tau_{22'}(0) = 33.5982645303$\,s $\pm 1.6$\,ns & $\delta\tau_{22'} = 5.2$\,ns \\
			~ & $2\beta_3 = +300.000013$\,ns/s $\pm 26$\,fs/s & $\tau_{33'}(0) = 33.2146434958$\,s $\pm 0.58$\,ns & $\delta\tau_{33'} = 1.9$\,ns \\
			\hline
			3 (TDI 2.0)& $2\beta_2 = -200.000028$\,ns/s $\pm 68$\,fs/s & $\tau_{22'}(0) = 33.5982645401$\,s $\pm 1.5$\,ns & $\delta\tau_{22'} = 5.0$\,ns \\
			~ & $2\beta_3 = +300.000020$\,ns/s $\pm 25$\,fs/s & $\tau_{33'}(0) = 33.2146435166$\,s $\pm 0.58$\,ns & $\delta\tau_{33'} = 1.9$\,ns \\			
			\hline
			~ & ~ & ~ & ~\\
			~~~One-Way Ranging & ~ & ~ & ~\\
			\hline
			\hline
			3 (TDI 2.0)& $\beta_2 ~ = -103.3$\,ns/s $\pm 1.4$\,ns/s  & $\tau_{2}(0)~~= 16.68021$\,s $\pm 32\,\mu$s & $\delta\tau_{2}~~= 105\,\mu$s \\
			~ & ~ & $\tau_{2'}^*(0)~= 16.91805$\,s $\pm 32\,\mu$s 	  & $\delta\tau_{2'}~= 105\,\mu$s \\
			~ & $\beta_3 ~ = +152.24$\,ns/s $\pm 1.4$\,ns/s    & $\tau_{3}(0)~~= 16.48824$\,s $\pm 31\,\mu$s & $\delta\tau_{3}~~= 99\,\mu$s \\
			~ & ~ & $\tau_{3'}^*(0)~= 16.72640$\,s $\pm 31\,\mu$s  & $\delta\tau_{3'}~= 99\,\mu$s \\
			\hline
			\hline
			~ & ~ & ~ & ~ \\
			Dynamic LISA-like & ~ & ~ & ~\\
			~with Confusion Noise & ~ & ~ & ~\\
			~ & ~ & ~ & ~\\
			~~~Round-trip Ranging & ~ & ~ & ~\\
			\hline
			\hline
			3 (TDI 2.0)& $2\beta_2 = -199.999991$\,ns/s $\pm 50$\,fs/s & $\tau_{22'}(0) = 33.6012734891$\,s $\pm 1.1$\,ns & $\delta\tau_{22'} = 3.7$\,ns \\
			~ & $2\beta_3 = +300.000137$\,ns/s $\pm 22$\,fs/s & $\tau_{33'}(0) = 33.2100302983$\,s $\pm 0.49$\,ns & $\delta\tau_{33'} = 1.6$\,ns \\		
			\hline
			~ & ~ & ~ & ~\\
			~~~One-Way Ranging & ~ & ~ & ~\\
			\hline
			\hline
			3 (TDI 2.0)~\ddag& $\beta_2 ~ = -96.81$\,ns/s $\pm 2.3$\,ns/s  & $\tau_{2}(0)~~= 16.73546$\,s $\pm 53\,\mu$s & $\delta\tau_{2}~~= 169\,\mu$s \\
			~ & ~ & $\tau_{2'}^*(0)~= 16.86582$\,s $\pm 53\,\mu$s  & $\delta\tau_{2'}~= 169\,\mu$s \\
			~ & $\beta_3 ~ = +149.439$\,ns/s $\pm 1.4$\,ns/s    & $\tau_{3}(0)~~= 16.53994$\,s $\pm 32\,\mu$s & $\delta\tau_{3}~~= 102\,\mu$s \\
			~ & ~ & $\tau_{3'}^*(0)~= 16.67009$\,s $\pm 32\,\mu$s  & $\delta\tau_{3'}~= 102\,\mu$s \\
			\end{tabular}
	\end{ruledtabular}
	\begin{flushleft}
	 \ddag~~The additional error in the one-way confusion noise measurement as compared to the phase-locked measurement is expected to be due to the coupling of low-frequency noise using the minimized-RMS ranging method \cite{TDIRanging}. \\
	~~\\
	*~~The returning delay times tend to be longer than the outgoing delay times by 100-300\,ms as a result of internal delays within the DSP system's EPD units.
	\end{flushleft}
\end{table*}

\subsection{Dynamic Arm Transponder \\ (TDI 2.0 Verification)}

In the next experiment, we incorporate time-dependent arm-lengths into the simulation with the $\beta$-values defined in Table\,\ref{tab:TDIMeasurements}. Again, initially assuming $\beta = 0$, the 40000\,s measurements of the $s_{21}$ and $s_{31}$ signals are broken into 40, 1000\,s segments. These data segments are then used to minimize the ranging tone and calculate the round-trip time-delay for each segment as defined by Fig.\,\ref{fig:FlowChart}. The linear regression of these time-delay offsets provides a first-order measure of the $\beta$ to an accuracy of $100 ps/s$ as shown in Table\,\ref{tab:RangingResults}. The process also calculates a first order measure of the round-trip time delay with a ranging precision of $< 7.5\,\mu$s ($\sim$1.7\,km) but, due to the incorrect $\beta=0$ assumption, these values tend to equal the average delay for the data-segment. A second iteration improves the $\beta$ accuracy to $80$\,fs/s and the ranging precision to $< 5.9$\,ns ($\sim$1.8\,m). The final iteration 
optimizes the $\beta$ precision to $8.9$\,fs/s and the ranging precision to $<0.65$\,ns ($\sim$0.2\,m). (Table\,\ref{tab:RangingResults}) Producing comparable values to within the measurement error, additional iterations do little to improve the tone cancellation or ranging accuracy.

\begin{figure}
\includegraphics[trim=2cm 0.5cm 3cm .5cm, clip=true, width=.48\textwidth]{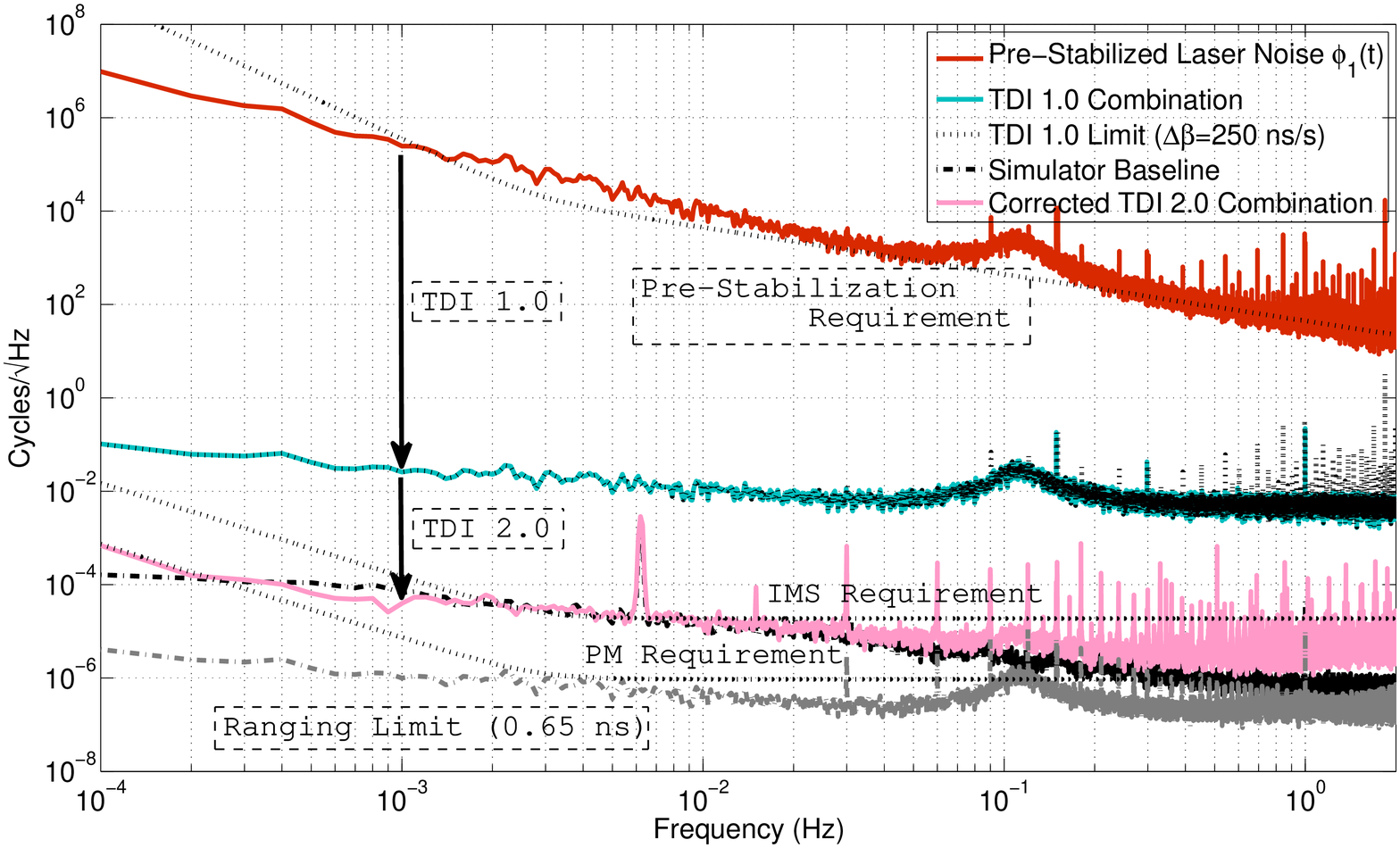}
\caption{\label{fig:TDI2Transponder} Dynamic Transponder Experimental Results: The suppression of the TDI $X_1$ combination is limited by the arm-length time-dependence. The TDI $X_2$ combination removes the additional time-dependent-coupled laser noise and reveals the 6.22 mHz GW signal. As with the static case (Fig.\,\ref{fig:TDI1Rescaled}), the EPD unit's phase transmission accuracy is the primary limiting noise source in the TDI combinations although, some sensitivity loss may occur due to a limited ranging capability around 100\,mHz (Fig.\,\ref{fig:XCorr}).}
\end{figure}

Applying the calculated round-trip functional values from the third iteration of the ranging procedure, we use the entire data-set to produce the TDI-$X_1$ and TDI-$X_2$ combinations (Fig.\,\ref{fig:TDI2Transponder}). The TDI-$X_1$ combination is limited, as theoretically anticipated, by Eq.\,\ref{eqn:TDICornish} with $\tau \simeq 16.7\,$s and $|\Delta\beta| = 250$\,ns/s. The TDI-$X_2$ combination's correction terms account for this dynamic arm-length limitation and remove the velocity dependent laser phase noise resulting in a sensitivity equal to the experiment's baseline noise performance. This result meets the IMS requirement defined by the LISA mission concept design. (Table\,\ref{tab:Requirements}) The ranging precision, as plotted in Fig.\,\ref{fig:TDI2Transponder}, is not expected to be a limiting noise source which is verified with through the cross-correlation of the input noise with the TDI-$X_2$ combination as shown in Fig.\,\ref{fig:XCorr}.

\subsection{Dynamic LISA-like \\ (Phase-locked Laser, One-Way Ranging)}
\label{sec:TDI2}

At this point, we include the phase-locking of the $L_{2/3}$ signals on the far spacecraft (Fig.\,\ref{fig:TDIExperiment}) and the transmission of these phase signals back to the local-SC, thus generating and measuring all four $s_{sr}$ beatnote observables. These signals are used in the same iterative process as previously described (Fig.\,\ref{fig:FlowChart}, Table\,\ref{tab:RangingResults}). The optimized time-delay functions from this process result in a measure of $|2\beta|$ to an accuracy better than $\sim$70\,fs/s and a round-trip ranging precision of $\sim$5.0\,ns ($\sim$1.5\,m). The constraints on the one-way delay times through the residual PLL noise removal ($\sim$1\,mHz/$\sqrt{\rm{Hz}}$) are not applicable until the precision of the round-trip delay times are accurate enough to remove enough of the input laser noise from the TDI combinations to reveal these residual PLL noises. Thus, it requires at least one iteration of the ranging process before one can constrain the one-way delay times. 
Applying a linear regression to the calculated one-way delay times we find a one-way ranging error of $\sim100\,\mu$s ($\sim$30\,km). The outgoing and return delay times are un-equal by $\sim250\pm 0.1\,$ms, proving the ability to extract the individual one-way laser phase errors despite un-equal delays along a single arm $(\tau_q(0) \neq \tau_{q'}(0))$.

\begin{figure}
\includegraphics[trim=2cm 0.5cm 3cm .5cm, clip=true, width=.48\textwidth]{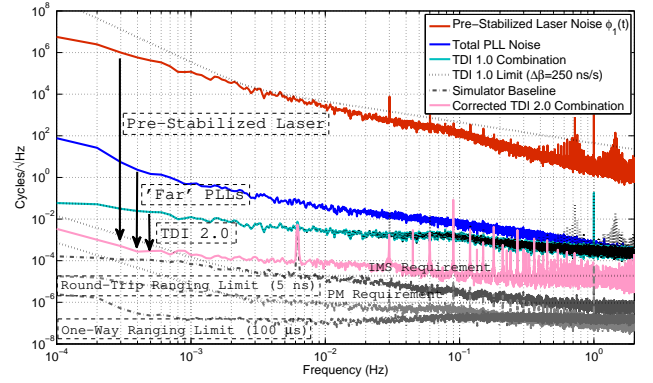}
\caption{\label{fig:TDI2PLL} Dynamic LISA-like (Phase-locked) Experimental Results: The suppression of the TDI $X_1$ combination is limited by the expected arm-length time-dependence. The TDI $X_2$ combination removes the input laser noise, the far-SC PLL residual phase noise, and the time-dependent coupled laser noise to reveal the 6.22 mHz GW signal. The sensitivity limitation comes, most likely, as a result of multiple uncorrelated EPD noise sources.}
\end{figure}

\begin{figure}
\includegraphics[trim=1.0cm 0cm 2cm .5cm, clip=true, width=.48\textwidth]{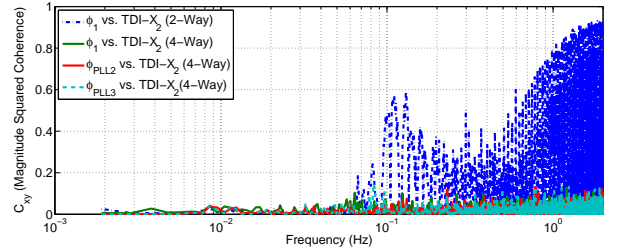}
\caption{\label{fig:XCorr} TDI vs. Input Cross Correlation: The magnitude squared cross-correlation of the LISA-like TDI measurements show no correlation with the input noise or either PLL noise source, verifying that all the laser noise sources have been sufficiently removed. The cross-correlation of the dynamic transponder TDI measurement shows some input phase correlated noise cancellation limitation for frequencies above 100\,mHz.}
\end{figure}

Applying these optimized one-way functional values from the ranging procedure, we produce the TDI-$X_1$ and TDI-$X_2$ combinations (Fig.\,\ref{fig:TDI2PLL}). Again, the TDI-$X_1$ combination equals the expected limitation (Eq.\,\ref{eqn:TDICornish}). The TDI-$X_2$ combination meets the LISA IMS requirement to within a factor of 4 and is likely limited by a combination of multiple EPD clock-noise sources resulting in a sensitivity greater than the simulator's baseline performance. Based on the variance of the fitted delay times, the ranging precisions are not a limiting noise source as plotted in Fig.\,\ref{fig:TDI2PLL}. The cross-correlation of the TDI-$X_2$ combination with the laser and PLL noise sources (Fig.\,\ref{fig:XCorr}) indicates that all the known and accounted-for noise sources have been sufficiently removed from the interferometer's output.

\subsubsection{Dynamic LISA-like \\(with Confusion Noise Background)}
\label{sec:TDIwCN}

Lastly, we include a low frequency simulated `confusion noise' into the measurement to ensure that these low-frequency terms do not limit the ranging precision. The confusion noise background, $\simeq 6.4/(f\sqrt{1+(f/f_{R})^2})\,\mu$cycles/$\sqrt{\rm{Hz}}$ where $f_{R}=1$\,mHz \cite{ConfusionEstimate1}, and the 6.22\,mHz mono-chromatic binary are simultaneously injected. The optimized ranging result places bounds on the $|2\beta|$ accuracy better than $\sim$50\,fs/s produces a round-trip ranging precision of $\sim$3.7\,ns ($\sim$1.1\,m). Thus, this confusion noise result achieves a ranging precision on the same order as the simulator's phase-locked performance, indicating that low-frequency noise has little to no effect on the ranging tone cancellation or the measured arm-lengths.

\begin{figure}
\includegraphics[trim=2cm 0.5cm 2cm .5cm, clip=true, width=.48\textwidth]{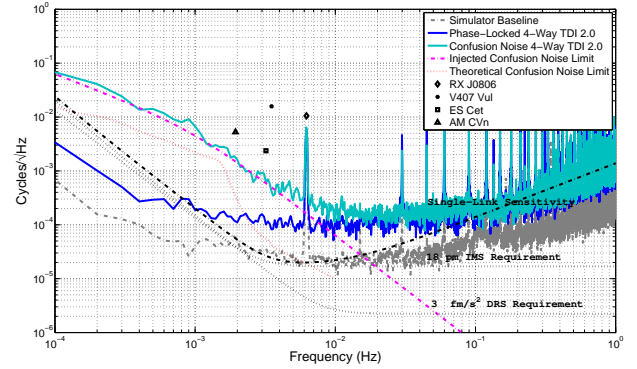}
\caption{\label{fig:TDIwConfusionNoise} Compiled Results and Comparison with TDI for LISA: In this figure we have compiled all the results of the TDI simulations and attempt to make a direct comparison with the expected LISA strain sensitivity.  The baseline spectral noise of the UFLIS simulator (grey-blue) from the TDI-Transponder measurements is plotted.  The velocity corrected TDI-$X_{2.0}$ spectrum of the dynamic arm TDI simulation with (cyan) and without (blue) the injected binary confusion noise (dotted-magenta) is plotted in comparison with the IMS sensitivity requirement.  The three TDI simulations are scaled to account for the high-frequency GW-sensitivity loss expected in LISA.  The DRS and IMS requirement are root-square summed and scaled by the high-frequency LISA GW-sensitivity loss function to produce the effective single-link LISA sensitivity.  An estimate of the confusion noise limit is plotted (dotted-red) along with the 
four brightest verification binaries rescaled from a 1-year averaged strain sensitivity to noise spectra in $cycles/\sqrt{Hz}$.  The strain magnitude of the $1\,year$ averaged RX-J0806 binary and the $10000\,s$ EPD injected GW have amplitudes such that they result in similar LSD amplitudes in this figure.}
\end{figure}

The noise spectrum comparisons between the TDI-$X_2$ outputs of the dynamic LISA-like experiments, with and without the confusion noise background, are plotted in Fig.\,\ref{fig:TDIwConfusionNoise} to show the additional low-frequency noise.  These spectra are a factor of 5 larger than the simulators baseline performance likely resulting from the coupling of the additional EPD, phasemeter, and electronic components in these experiments.  All three measured spectra in Fig.\,\ref{fig:TDIwConfusionNoise} have been scaled by the high-frequency sensitivity loss `roll-up' of the LISA-detector for GW-frequencies larger than $1/\tau = 60$\,mHz in order to obtain a direct comparison with the single-link sensitivity. The injected confusion noise background is plotted and matches the spectrum of measured confusion noise.  A theoretical confusion-noise background \cite{ConfusionEstimate1} and the expected 1-year strain amplitude of the four strongest LISA verification binaries \cite{LISAVerBinary} are marked for 
additional reference.

\begin{figure}
\includegraphics[trim=2cm 0cm 2cm 0cm, clip=true, width=.48\textwidth]{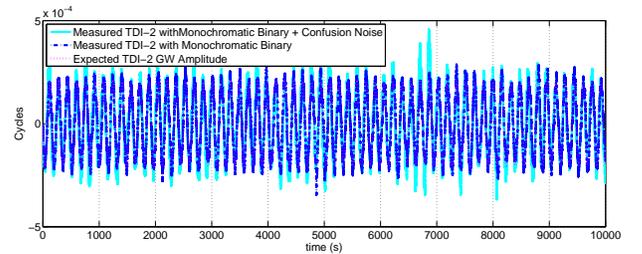}
\caption{\label{fig:TDIwCNTimeSeries} TDI-$X_2$ Time-Series: The TDI-$X_2$ extracted signals match the expected GW amplitude given the $51.2\,\mu$cycle GW EPD injection amplitude after scaling by the magnitude of the TDI-$X_2$ combination's transfer function evaluated at $f = 6.22\,$mHz, $|\tilde{X}_{2}(6.22\,\rm{mHz})| = 2.33$.}
\end{figure}

The time-series of the extracted monochromatic GW binary signal using the TDI-$X_2$ combination is plotted in Fig.\,\ref{fig:TDIwCNTimeSeries} with and without the confusion noise background. Comparing the measured amplitude with the expected GW amplitude, given the injected $51.2\,\mu$cycle EPD GW-signal, we find the TDI-$X_2$ extracted signal matches the expected amplitude of $4\times51.2\,\mu$cycles$ = 205\,\mu$cycles.  The factor of 2 accounts for the two interferometer arms while the factor of 2.33 accounts for the TDI transfer function's signal gain at $f = 6.22\,$mHz ($|\tilde{X}_{2}(6.22\,\rm{mHz})| = 2.33$).

\section{\label{sec:Conc}Conclusion}

Following the initial interferometry tests of a static LISA model \cite{UFLIStdi1}, we expanded UFLIS and added time varying signal travel times, Doppler shifts, and gravitational wave signals to our electronic phase delay units. This enables tests of the LISA interferometry in a realistic, dynamic model. Our experimental results show that than 10 orders of magnitude of laser phase noise can be canceled using appropriately time-shifted data streams in the TDI-$X_2$ data combination. We also developed and demonstrated a simple but powerful ranging method to measure the signal travel times between the spacecraft.

We verified that the ability to reduce laser phase noise using a TDI-$X_1$ data combination is indeed limited by the relative velocities between the spacecraft. Furthermore, we demonstrated the removal of the residual phase lock loop noise added at the far spacecraft and, in this configuration, showed that the requirements on one-way ranging are relaxed by several orders of magnitude compared to the requirements on round trip ranging.

In the process, we developed and tested data analysis tools which take the raw phasemeter data streams, extracts the light-travel time functions, and generates the TDI-$X_2$ data sets. We also added a confusion noise GW-background to our interferometry emulator and verified that this background noise does not interfere with our ranging capabilities.

Future experiments should include real, LISA-like GW signals using data-sets generated with LISA-tools like Synthetic LISA \cite{SynthLISA}. Simulations with three independently stabilized lasers might also be valuable towards verifying the constrains on the one-way ranging capabilities.

\bibliography{TDI2}

\end{document}